\documentclass[prl,reprint,aps,amsmath,amssymb,twocolumn,superscriptaddress,bibliography]{revtex4-1}
\usepackage{CJK}
\pdfoutput=1
\usepackage{graphicx}
\usepackage{caption}
\usepackage{overpic}
\usepackage{dcolumn}
\usepackage{bm}
\usepackage{subfigure}
\usepackage{amsmath}
\usepackage{amsfonts}
\usepackage{color}
\usepackage{array} 
\usepackage{booktabs} %
\usepackage{rotating}
\usepackage[english]{babel}
\usepackage{amsthm,amsmath,amssymb}
\usepackage{mathrsfs}
\usepackage{lipsum}
\usepackage{mathptmx}
\DeclareMathAlphabet{\mathcal}{OMS}{cmsy}{m}{n}
\DeclareSymbolFont{largesymbols}{OMX}{cmex}{m}{n}
\let\sum\relax
\DeclareSymbolFont{CMlargesymbols}{OMX}{cmex}{m}{n}
\DeclareMathSymbol{\sum}{\mathop}{CMlargesymbols}{"50}

\usepackage{booktabs,array,float,tabularx,booktabs,lipsum,amsmath,multirow}
\usepackage[hidelinks]{hyperref}
\hypersetup{
    colorlinks=true,
    linkcolor=blue,
    filecolor=magenta,
    urlcolor=blue,
    citecolor=magenta,
    }   

\usepackage{float}
\DeclareMathOperator{\sech}{sech}
\usepackage[
text={7in,10in},centering,
height=10in,a4paper,hmargin={1.5cm,1.5cm},
]{geometry}

\begin{document}

\title{Discovery and regulation of chiral magnetic solitons: Exact solution from Landau-Lifshitz-Gilbert equation}
\author{Xin-Wei Jin}
\affiliation{School of Physics, Northwest University, Xi'an 710127, China}
\affiliation{Peng Huanwu Center for Fundamental Theory, Xi'an 710127, China}

\author{Zhan-Ying Yang}
\email{zyyang@nwu.edu.cn}
\affiliation{School of Physics, Northwest University, Xi'an 710127, China}
\affiliation{Peng Huanwu Center for Fundamental Theory, Xi'an 710127, China}

\author{Zhimin Liao}
\affiliation{School of Physics, Peking University, Beijing, 100871,China}

\author{Guangyin Jing}
\email{jing@nwu.edu.cn}
\affiliation{School of Physics, Northwest University, Xi'an 710127, China}

\author{Wen-Li Yang}
\affiliation{Peng Huanwu Center for Fundamental Theory, Xi'an 710127, China}
\affiliation{Insititute of Physics, Northwest University, Xi'an 710127, China}

\date{\today}

\begin{abstract}
The Landau-Lifshitz-Gilbert (LLG) equation has emerged as a fundamental and indispensable framework within the realm of magnetism. However, solving the LLG equation, encompassing full nonlinearity amidst intricate complexities, presents formidable challenges. Here, we develop a precise mapping through geometric representation, establishing a direct linkage between the LLG equation and an integrable generalized nonlinear Schr\"odinger equation. This novel mapping provides accessibility towards acquiring a great number of exact spatiotemporal solutions. Notably, exact chiral magnetic solitons, critical for stability and controllability in propagation with and without damping effects are discovered. Our formulation provides exact solutions for the long-standing fully nonlinear problem, facilitating practical control through spin current injection in magnetic memory applications.
\end{abstract}

\maketitle

\textit{Introduction.\textemdash}
The seminal 1935 work by Landau and Lifshitz, which laid down the foundational dynamical equation governing magnetization based on phenomenological insights \cite{landau1992theory,saslow2009landau,lakshmanan2011fascinating}, and the subsequent introduction of a damping term by Gilbert \cite{gilbert2004phenomenological}, the amalgamation of these concepts has given rise to the renowned Landau-Lifshitz-Gilbert (LLG) equation. Over the years, this equation has emerged as a fundamental and indispensable framework within magnetism field. Its contemporary significance has been amplified through remarkable advancements, most notably the incorporation of an additional term that facilitates the explication of spin torque phenomena in spintronics \cite{slonczewski1996current,apalkov2013spin,li2004domain2,liu2020three,zhang2004roles,yang2015domain}, spin waves \cite{ahlberg2022freezing,yazdi2021tuning,wang2023long,liu2018long,li2021topological,lan2021geometric,pribiag2007magnetic,yu2021magnetic}, magnetic solitons \cite{kamenetskii2021chirality,ohkuma2020soliton,parkin2008magnetic,kosevich1990magnetic,zhang2020skyrmion,muhlbauer2009skyrmion,zhang2016magnetic,dohi2019formation,togawa2012chiral,osorio2021creation}, spatio-temporal patterns \cite{goussev2010domain,iacocca2017breaking}, and even chaotic behavior \cite{yang2007chaotic}. Further advancements have paved the way for applications in next-generation magnetic storage \cite{gu2022three,zhang2017stateful,siracusano2016magnetic}, neural networks \cite{torrejon2017neuromorphic,yang2022spintronic,wang2023spintronic}, and logic gates \cite{luo2020current,manipatruni2019scalable,wang2022ferroelectric,shen2023programmable,allwood2005magnetic}.

Despite its deceptively simple appearance, solving the LLG equation poses an exceptional challenge \cite{goussev2010domain,iacocca2017breaking}, rendering it a persistently unresolved problem for nearly nine decades. This complexity emanates from its intricate nature as a vector-based highly nonlinear partial differential equation. In real-world scenarios, the LLG equation encompasses a myriad of complex interactions among the components of the magnetization vector \cite{gilbert2004phenomenological}. Consequently, solutions often necessitate recourse to linearization, approximations, and asymptotic techniques such as the Holstein-Primakoff (HP) transformation \cite{holstein1940field,daniel2008soliton}, reductive perturbation scheme \cite{mikeska1991solitary,iacocca2017symmetry}, and long wavelength approximation. Nonetheless, these techniques prove utterly ineffectual in regions of large amplitudes or strong nonlinearity.
Therefore, exact solution of the LLG equation emerges as a potent bridge, overcoming these gaps and revealing profound revelations regarding magnetization dynamics, thereby furnishing insightful understandings for simulating and comprehending intricate magnetic systems.

In this Letter, through a geometric representation \cite{lakshmanan1976dynamics}, we establish an exact mapping of the LLG equation onto an integrable generalized nonlinear Schr\"odinger equation, free of any approximation.
This novel mapping provides accessibility towards acquiring a great number of exact spatiotemporal solutions of the original equation. Notably, we unveil an analytical formulation for chiral magnetic solitons, encompassing a spectrum ranging from left-handed, neutral to right-handed configurations, determined by a defined chirality factor.
The derived exact solution indicates the potential for arbitrary manipulation of magnetic soliton motion through the injection of spin current
\textemdash
a discovery that aligns seamlessly with our numerical findings. To encapsulate the realism of dissipative devices, we incorporate Gilbert damping into the dynamics of these chiral magnetic solitons, thereby estimating their dynamic propagation.

\begin{figure}[htbp]
\vspace{0cm} %
\centering
\includegraphics[width=8.5cm]{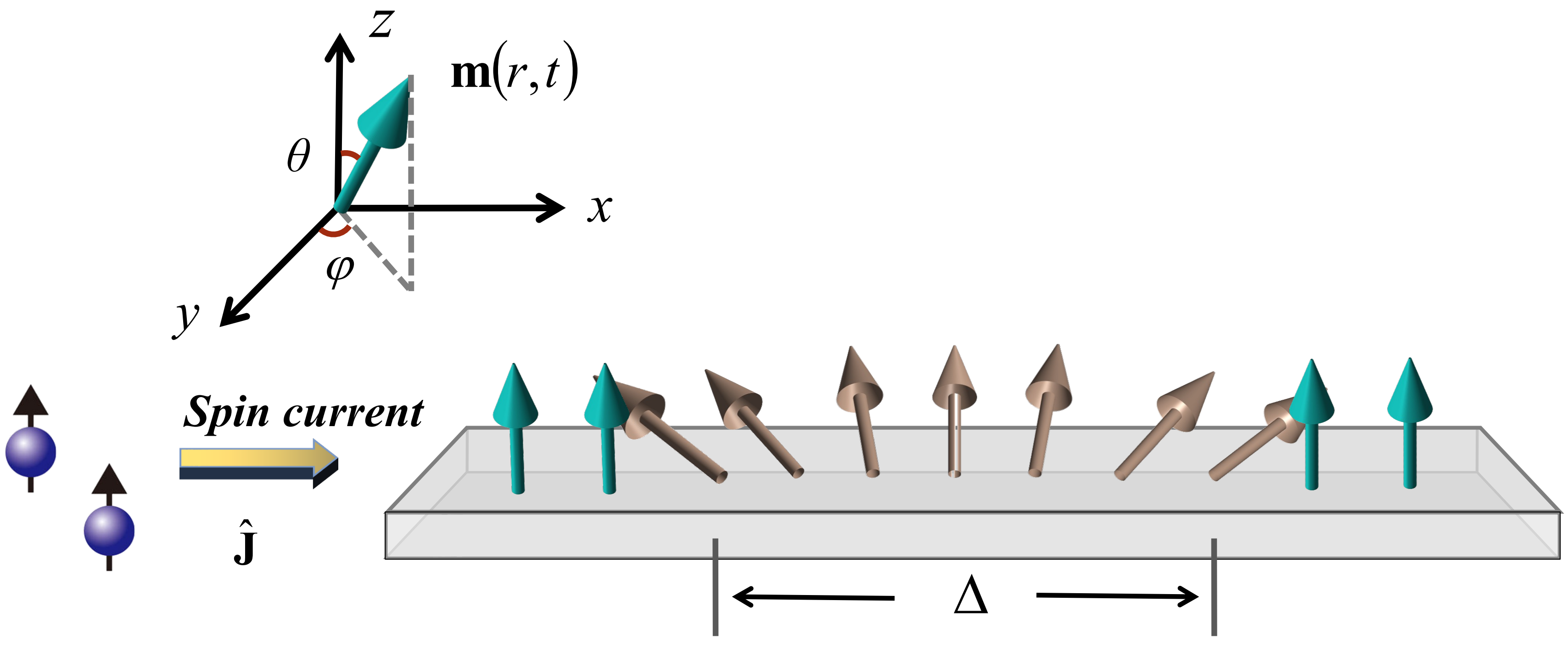}
\caption{Schematic diagram of 1D ferromagnetic structure. Magnetic soliton excitation driven by spin-polarized currents. Here $\Delta$ represents the width of magnetic solitons.}\label{Sketch}
\end{figure}

\textit{Modeling.\textemdash}We consider an isotropic ferromagnetic nanowire with spin-polarized current flowing along the axis of nanowire as depicted in Fig. \ref{Sketch}.
A ``nanowire" as defined here is a planar ferromagnetic stripe of length $L_{x}$, width $L_{y}$, and thickness $L_{z}$ along $\hat{\textbf{x}}$ , $\hat{\textbf{y}}$, and $\hat{\textbf{z}}$, respectively, with $L_{x}\gg L_{y}>L_{z}$.

\begin{figure*}[htbp]
\vspace{0cm} %
\centering
\includegraphics[width=12cm]{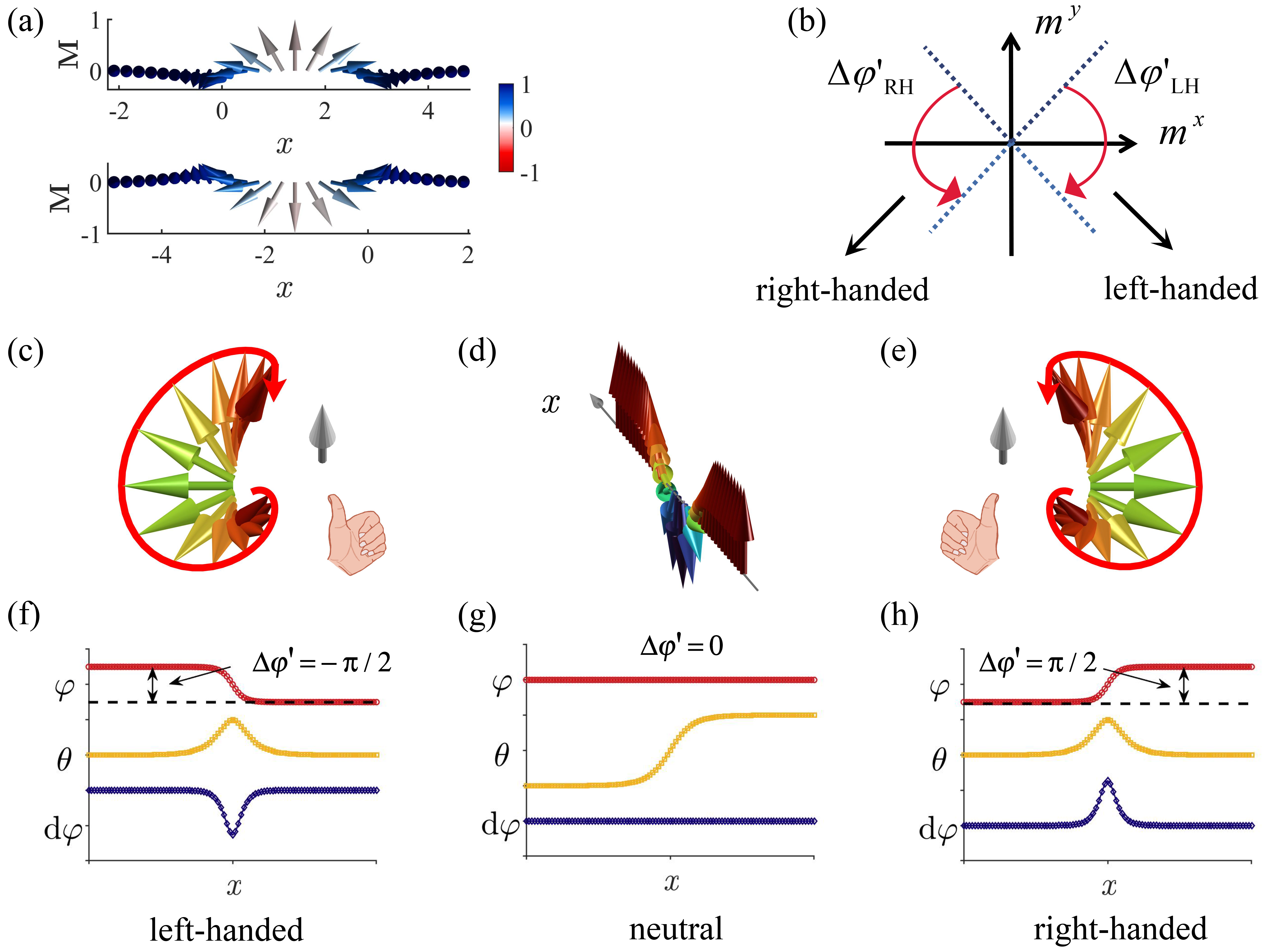}
\caption{Spatial structure and classification of chiral magnetic solitons. (a) Vertical views of the left-handed and right-handed magnetic solitons. (b) Schematic plot of the chirality defined by the azimuth angle change. The pair of red arrows delineate the azimuthal directional changes of the left and right chiral magnetic solitons across the distribution axis. Their discrepancies in azimuthal variation are denoted by $\Delta\varphi'_{LH}$ and $\Delta\varphi'_{RH}$. (c)-(e) Spatial spin structures of left-handed magnetic soliton, neutral magnetic soliton, and right-handed magnetic soliton. (f)-(h) illustrate the azimuthal,
polar angle, and phase gradient flow of the three kinds of chiral solitons.}\label{Solution}
\end{figure*}

The magnetization dynamics is described by the famous LLG equation
\begin{equation}\label{LLG}
\frac{\partial\textbf{m}}{\partial t}=-\gamma\textbf{m}\times\textbf{H}_{\rm{eff}}+\alpha\left(\textbf{m}\times\frac{\partial\textbf{m}}{\partial t}\right)+{\rm\bm{\tau}}_{b},
\end{equation}
where $\textbf{m} = \textbf{M}/M_{s}=(m^{x},m^{y},m^{z})$ is the unit magnetization vector with $M_{s}$ being the saturated magnetization.
The first term on the right-hand side represents the torque contributed by the effective field $\textbf{H}_{\rm{eff}}$ (including applied, demagnetizing, anisotropy, and exchange fields), $\gamma$ is the gyromagnetic constant.
The second term describes the Gilbert damping torque, parameterized by a dimensionless damping factor $\alpha$.
The last term ${\rm\bm{\tau}}_{b}$ represents the spin-transfer torque (STT), which comprises dual components that can be written as ${\rm\bm{\tau}}_{b}=-b_{J}(\hat{\textbf{J}}\cdot\nabla)\textbf{M}+\beta b_{J}\textbf{M}\times(\hat{\textbf{J}}\cdot\nabla)\textbf{M}$. Here $\hat{\textbf{J}}$ is the unit vector in the direction of the current.
These two components are most commonly termed adiabatic and non-adiabatic spin torques, respectively, with $b_{J}=Pj_{e}\mu_{B}/(eM_{s})$ and $\beta$ defined as the non-adiabatic torque coefficient.
Wherein, $P$ represents the spin polarization of current,
$j_{e}$ is the electric current density, $\mu_{B}$ is the Bohr magneton, and $e$ is the magnitude of electron charge.
In what follows, we take only adiabatic STT into consideration for two reasons:
one is that the most widely agreed upon interaction between a spin-polarized current and a magnetic soliton is adiabatic STT; and the other is that the magnitude of the nonadiabatic spin torque is about 2 orders of magnitude smaller than adiabatic torque $(\beta\approx10^{-2})$.
Let us begin by examining the most elementary effective field, encompassing solely exchange fields, i.e. $\textbf{H}_{\rm{eff}}=(2\mathcal{A}/M_{s})\nabla^{2}\textbf{m}$, where $\mathcal{A}$ is the exchange stiffness constant.

The spatiotemporal transformation $\tau=\gamma\mu_{0}M_{s} t/(1+\alpha^2)$ and $\zeta=\lambda_{ex}\cdot x$ are introduced to recast the LLG equation into the dimensionless Landau-Lifshitz form (Note that $\lambda_{ex}=\sqrt{2A/(\mu_{0}M_{s}^{2})}$ is the exchange length):
\begin{equation}\label{rLLG}
\begin{aligned}
\textbf{m}_\tau=-\textbf{m}\times\textbf{m}_{\zeta\zeta}
-\alpha\textbf{m}\times\left(\textbf{m}\times\textbf{m}_{\zeta\zeta}\right)+\mathbb{Q}\textbf{m}_{\zeta},
\end{aligned}
\end{equation}
where $\mathbb{Q}=b_{J}(1+\alpha\beta)/\sqrt{2\mathcal{A}\gamma^{2}\mu_{0}}$, a dimensionless number measuring the ratio of external spin current over exchange interaction strength.
This dimensionless STT-LLG model (\ref{rLLG}) effectively describes the dynamics of nonlinear excitations, such as magnetic solitons, occurring in ferromagnetic nanowires upon spin injection. Moreover, it exhibits qualitative reproduction much of the behavior seen experimentally. For a permalloy nanowire, the standard material parameters are: $\gamma=1.76\times10^{11} {\rm rad/s\cdot T}, M_{s}=8\times10^{5} {\rm A/m}, \mathcal{A}=1.3\times10^{-11} {\rm J/m}, P=0.5$. As a result, the units in time and space after rescaling are $1 \tau\approx5.70\ {\rm ps}, 1 \zeta\approx5.68\ {\rm nm}$.

\textit{Chiral magnetic soliton.\textemdash}The high nonlinearity of STT-LLG model (\ref{rLLG}) presents a great challenge for comprehensive analytical research and restricts the exploration of novel spin textures to the realm of micromagnetic simulation or weak nonlinearity.
In this context, to obtain an analytical depiction of large-amplitude magnetic textures, we exactly map the STT-LLG equation (\ref{rLLG}) into a generalized nonlinear Schr\"odinger (GNLS) equation (See \textit{Supplementary Material} for further details on the spatial curve mapping
procedure), devoid of any approximations. For no damping, the GNLS equation reads
\begin{equation}\label{GNLS}
i\Psi_{\tau}+\Psi_{\zeta\zeta}+2|\Psi|^{2}\Psi-i \mathbb{Q} \Psi_{\zeta}=0.
\end{equation}

Exact cycloidal chiral magnetic soliton solutions can be constructed by applying the Darboux transformation (DT) \cite{baronio2012solutions,ling2016darboux}. Indeed, using the mapping relationship between equation solutions, the specific expression of three components of magnetization are obtained (See \textit{Supplementary Material} for detailed calculations):
\begin{equation}\label{One}
\begin{split}
m^{x}&=\frac{2a}{a^{2}+b^{2}}\left[a\sinh(\Xi)\sin(\Gamma)-b\cosh(\Xi)\cos(\Gamma)\right]\cdot\sech^{2}(\Xi),\\
m^{y}&=\frac{2a}{a^{2}+b^{2}}\left[a\sinh(\Xi)\cos(\Gamma)+b\cosh(\Xi)\sin(\Gamma)\right]\cdot\sech^{2}(\Xi),\\
m^{z}&=1-\frac{2a^{2}}{a^{2}+b^{2}}\sech^{2}(\Xi),\\
\end{split}
\end{equation}
with
\begin{equation*}
\begin{split}
\Xi&=2a\left[x+\int(\mathbb{Q}+4 b){\rm d}t\right],\\
\Gamma&=2b\left[x+\int(\mathbb{Q}-2(a^{2}-b^{2})/b){\rm d}t\right],
\end{split}
\end{equation*}
where $a$ and $b$ describe the wave number and the velocity of the magnetic soliton.

Fig. \ref{Solution}(a) depicts the spin textures of the obtained magnetic solitons.
Evidently, these solitons showcase a mirror symmetry relative to the wave vector axis, indicating their inherent chirality.
The chirality can be characterized by variations of the azimuth angle.
To clarify, we denote the polar and azimuthal angles of $\textbf{m}$ by $\theta$ and $\varphi$, respectively (as shown in Fig. 1), such that $M_{+}=m^{x}+im^{y}=M_{0}\sin(\theta)\exp(i\varphi), m^{z}=M_{0}\cos(\theta)$.
Note that the azimuthal angle exhibits a periodic background, which arises from the variation of magnetization in space, as revealed by the solution (\ref{One}).
The oscillation structures present in the azimuthal angle profiles are related to the small oscillation of $m^{\nu}$ ($\nu=x,y$), even though they may not be readily visible in Fig. \ref{Solution}(a).
By eliminating the meaningless periodic background phase, the real phase jumps of the chiral magnetic solitons are obtained by calculating the intrinsic argument $\varphi'(x) = \arg M'_{+}$, where $M'_{+} = M_{+} \exp(i\Gamma)$.
As a result, the azimuthal angles of both chiral solitons are demonstrated in Fig. \ref{Solution}(b).
Two red arrows span between the blue dashed line representing negative infinity and the corresponding positive infinity, delineating the azimuthal evolution of magnetic solitons across the distribution axis.
The distinction in azimuthal variation for chiral magnetic solitons are denoted as $\Delta\varphi'_{LH}$ and $\Delta\varphi'_{RH}$.
Notably, the two classes of chiral magnetic solitons exhibit opposite phase jumps, corresponding to two distinct chiralities.

The total phase change is defined as $\Delta\varphi'=\varphi'(x\to +\infty)-\varphi'(x\to -\infty)$. In general, the phase change of arbitrary magnetic solitons can be determined by integrating the phase gradient flow.
Insight can be gained from combining both argument and the phase gradient flow $\nabla\varphi'(x)$. Starting from $M'_{+}$ that constructed from exact solutions, we obtain
\begin{equation}
\nabla\varphi'(x)=\frac{2b\left[\sech(2\Xi)+1\right]}{\left(\frac{a^{2}-b^{2}}{a^{2}+b^{2}}\right)\sech(2\Xi)-1}.
\end{equation}
One can observe that the denominator of the aforementioned expression is consistently a non-positive value, which indicates that ``$+$" and ``$-$" families of phase gradient flow are characterized by the opposite signs of $b$.
Here, we define a chirality factor $\mathcal{C}={\rm sgn}(b)=\pm1$, which determines the chirality of magnetic solitons.
It is straightforward to verify that the nonzero phase variation is characterized by a simple expression:
$\Delta\varphi'=2\,\mathcal{C}\arctan(\lvert a/b \rvert).$
Thus, the chirality of the chiral magnetic soliton is entirely determined by this chirality factor $\mathcal{C}$.
When $b=0$, a special case naturally occurs, where the chirality factor cannot be defined, and chirality disappears, corresponding to the neutral cycloidal magnetic solitons.
Finally, we can now classify the exact solution (\ref{One}) into three categories based on the chirality factor, corresponding to neutral, left-handed, and right-handed chiral magnetic solitons. Figs. \ref{Solution}(c)\textendash\ref{Solution}(h) depict the typical spin textures, azimuthal angles, polar angles, and phase gradient flow at $t=0$ when no spin current is applied.

\begin{figure}[tbp]
\vspace{0cm} %
\centering
\includegraphics[width=8.5cm]{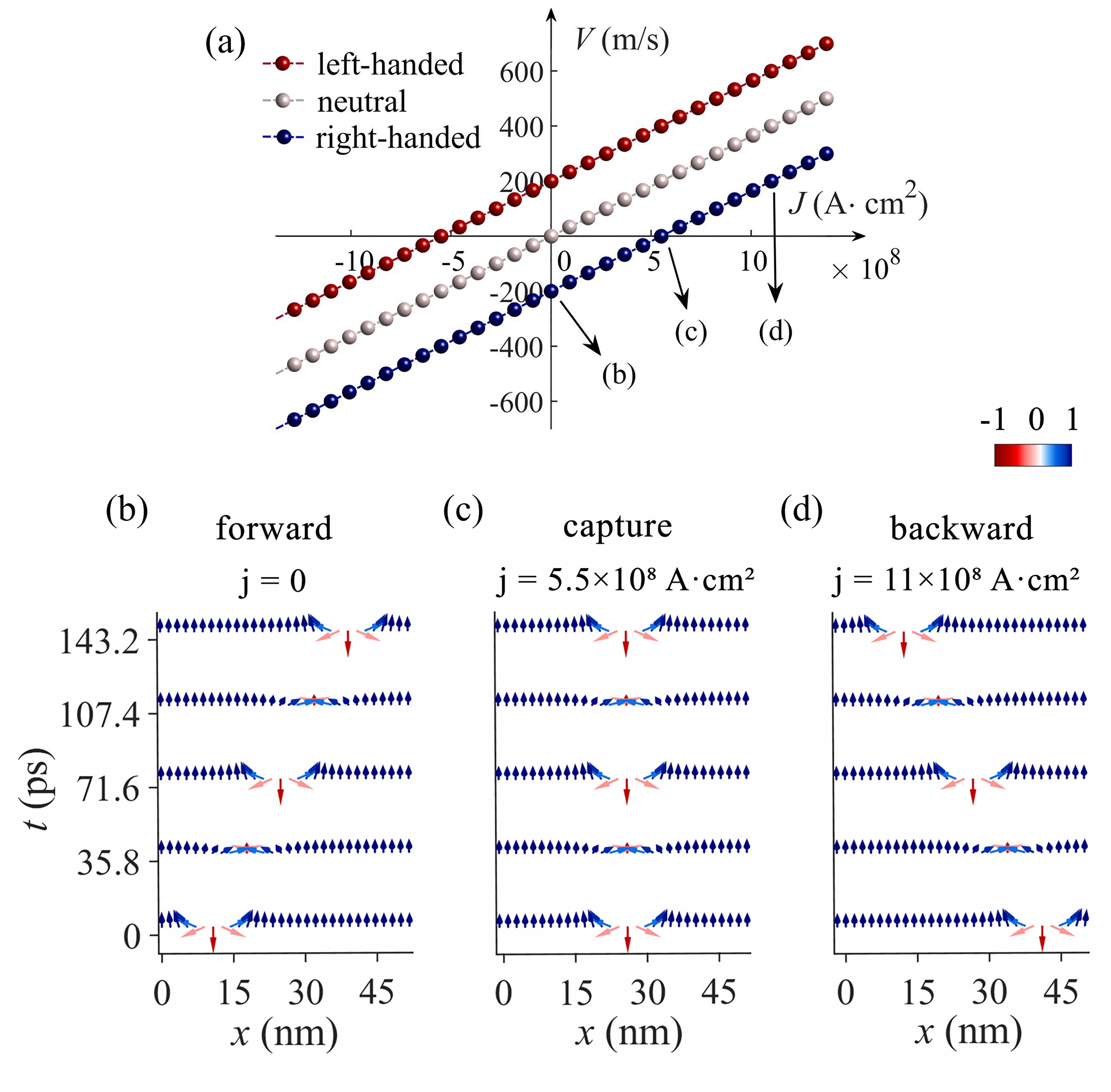}
\caption{ Coupling between chiral magnetic solitons and spin current injection. (a) Velocities of three distinct classes of chiral magnetic solitons plotted against spin-polarized currents ($j$). (b)-(d) Controlled manipulation of right-handed magnetic solitons under varying current strengths, enabling forward, backward, and arrested motion.}\label{STT}
\end{figure}

\begin{figure*}[htbp]
\vspace{0cm} %
\centering
\includegraphics[width=14cm]{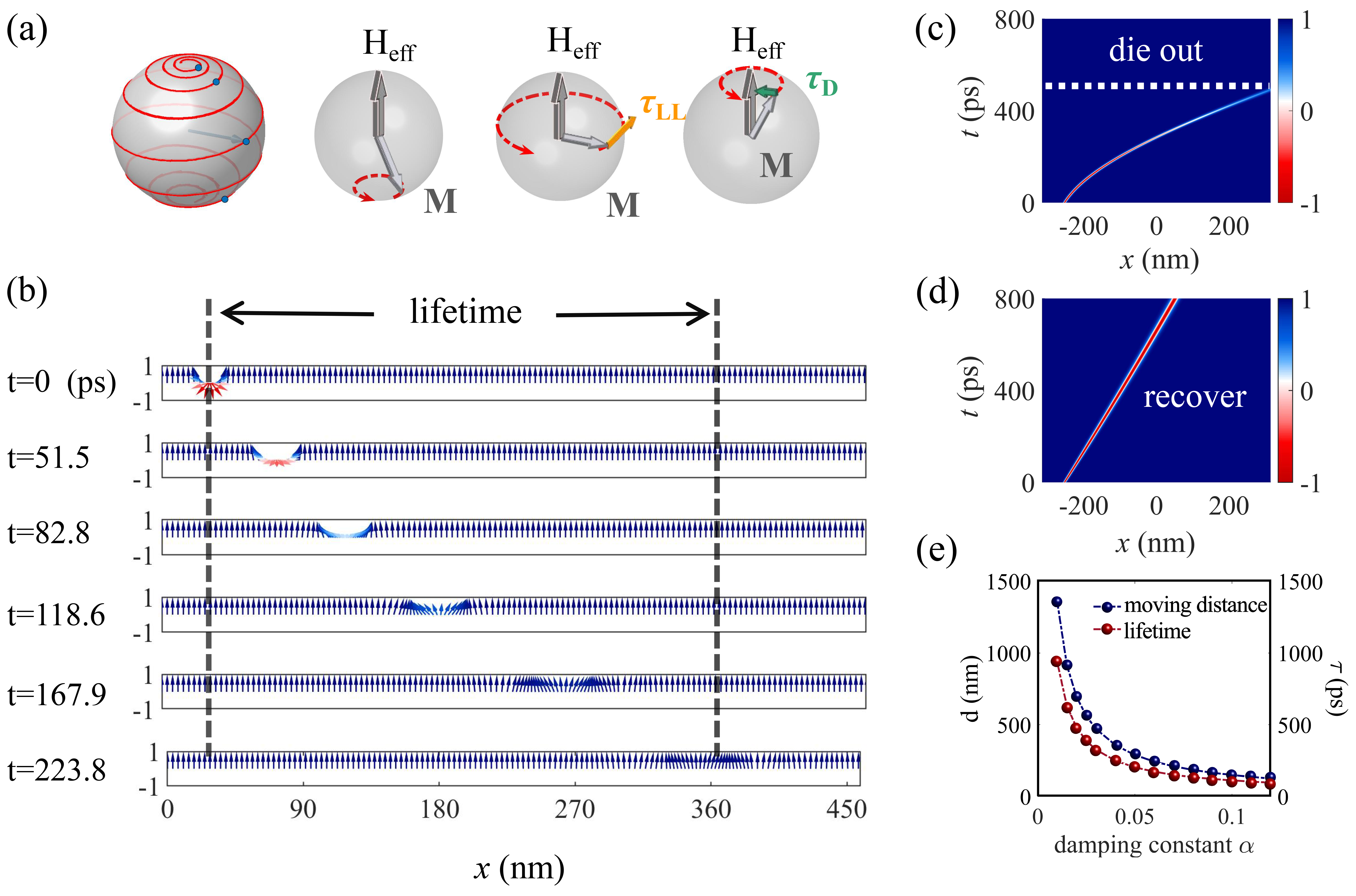}
\caption{Transmission of magnetic solitons in damped ferromagnetic nanowires and the anti-damping effect of non-adiabatic STT. (a) Schematic diagram of magnetization dissipation under damping. (b) Propagation of right-handed magnetic solitons in ferromagnetic nanowires with Gilbert-damping constant $\alpha=0.05$. (c) Temporal evolution of the magnetization component $m^{z}$ in the absence of non-adiabatic STT, where damping constant $\alpha=0.01$. (d) Temporal evolution of the magnetization component $m^{z}$ with spin-polarized current $j=3.7\times10^{7}{\rm A\cdot cm^{-2}}$. (e) Gilbert-damping dependence of lifetime and moving distance.} \label{Damping}
\end{figure*}

\textit{Spin-current coupling and damping effect.\textemdash}We now move to study the coupling between chiral magnetic soliton and the injection of spin current.
It has been demonstrated that the spin transfer torque is capable of driving the domain wall or skyrmion \cite{li2004domain2,dohi2019formation}, eliciting their prompt movement at a considerable velocity upon the application of spin current.
Here we report a comparable phenomenon on the chiral magnetic soliton from both theoretical and simulation results.
The numerical simulation results depicting the relationship between the velocities of three categories of chiral magnetic solitons and the injected spin currents are illustrated in Fig. \ref{STT}(a), and are in direct agreement with those obtained from the analytical solutions (\ref{One}).
These linear correlations can be realized from the equivalent GNLS equation (\ref{GNLS}), wherein the spin current term can be normalized to resemble the ``driving velocity", as supported by the dimensional analysis of $\mathbb{Q}$.
Figs. \ref{STT}(b)\textendash \ref{STT}(d) exemplify the current manipulation for right-handed magnetic soliton that comprise a series of transient snapshots captured during the magnetization evolution process (the model parameters are shown in the caption).
The above results highlight two notable aspects. Firstly, the chiral magnetic soliton possesses an inherent velocity linked to its initial magnetization state. Secondly, the solitons' motion can be stimulated by a spin-polarized current, while preserving their chirality. The external injection of spin current offers a means to manipulate chiral magnetic solitons, granting control over their forward, backward, and frozen motion.

Until now, our analysis is based on a perfect ferromagnetic wire in the absence of damping. Strictly speaking, in realistic nanowires, magnetic solitons cannot move over a large distance due to Gilbert damping.
The existence of damping introduces a small torque field, which dissipates the energy of the system during magnetization dynamics, and leads to a helical precession of the magnetization towards the direction of the effective field, i.e. the minimum energy state (See Fig. \ref{Damping}(a)).
To understand this damping effect in greater detail we performed numerical simulations of single-soliton dynamics.
Fig. \ref{Damping}(b) shows the evolution of a right-handed magnetic soliton in a nanowire with Gilbert-damping constant $\alpha = 0.05$.
It can be seen that the chiral magnetic soliton degenerates to a homogeneous magnetized state after propagating for about 223.8 $\rm{ps}$.
During the whole process, the magnetic soliton undergoes continuous deformation.
This has two consequences: magnetic soliton spreading and slowing of internal oscillations.
In order to characterize the presence of a chiral magnetic soliton, we define the soliton polarization $P_{s}=\frac{\left[1-\min(m_{z})\right]}{2}$.
The soliton is deemed to have dissipated when its polarization is lower than $5\%$ in comparison to the maximum magnetization.
The movement of magnetic solitons within ferromagnetic nanowires, subject to varying damping, results in distinct lifetimes. Fig. \ref{Damping}(e) depicts line graphs illustrating the relationship between the damping coefficient, lifetime, and moving distance.
The dissipation of solitons due to damping is a challenge to circumvent, and one approach is to seek ferromagnetic materials with low damping coefficients. Here, we explored the potential anti-damping effect of non-adiabatic STT, as depicted in Figs. \ref{Damping}(c) and \ref{Damping}(d). In the absence of external spin current, the $m^{z}$ component of the magnetic soliton diminishes during transmission. However, upon injecting an appropriate spin current, the incorporation of non-adiabatic STT enables the chiral magnetic soliton to propagate uniformly in its original velocity, resulting in a significantly extended lifetime.

\textit{Conclusions.\textemdash}
In this Letter, we have shown that the dimensionless LLG equation containing STT is entirely equivalent to the generalized nonlinear Schr\"odinger equation without any approximation.
This remarkable integrable system enables us to predict novel exact spatiotemporal magnetic solitons.
By applying the Darboux transformation, we obtain exact solution of chiral magnetic solitons, emerging within an isotropic ferromagnetic nanowire.
Our analytical formulation establishes a distinct correlation between chiral magnetic solitons and the infusion of spin currents, corroborating our numerical findings. This interrelation underscores the potential for arbitrary manipulation of magnetic soliton motion through spin current injection.
The inherent chirality of the micromagnetic structure plays a pivotal role in soliton motion: a reversal in chirality leads to a shift in motion direction.
To encapsulate the realism of dissipative devices, we investigate the influence of Gilbert damping on the motion of chiral magnetic solitons. The results reveal that in the presence of damping, chiral magnetic solitons gradually evolve toward a uniformly magnetized state. This implies a topological equivalence between the two magnetization states.
We propose selecting an appropriate spin current intensity to introduce an anti-damping effect, thereby ensuring the long-distance transmission of the chiral magnetic soliton. These results present new possibilities for developing chiral magnetic soliton-based racetrack memory.

\hspace*{\fill}

The authors thank Prof. H. M. Yu, Prof. L. C. Zhao, Prof. J. Liu and  Prof. C. P. Liu for their helpful discussions. This work was supported by the National Natural Science Foundation of China (No.12275213, 12174306,12247103), and Natural Science Basic Research Program of Shaanxi (2023-JC-JQ-02, 2021JCW-19).

\bibliography{references}
\end{document}


\title{\underline{Supplementary Materials} for ``Discovery and regulation of chiral magnetic solitons: Exact solution from Landau-Lifshitz-Gilbert equation"}

\author{Xin-Wei Jin}
\affiliation{School of Physics, Northwest University, Xi'an 710127, China}
\affiliation{Peng Huanwu Center for Fundamental Theory, Xi'an 710127, China}

\author{Zhan-Ying Yang}
\email{zyyang@nwu.edu.cn}
\affiliation{School of Physics, Northwest University, Xi'an 710127, China}
\affiliation{Peng Huanwu Center for Fundamental Theory, Xi'an 710127, China}

\author{Zhimin Liao}
\affiliation{School of Physics, Peking University, Beijing, 100871,China}

\author{Guangyin Jing}
\email{jing@nwu.edu.cn}
\affiliation{School of Physics, Northwest University, Xi'an 710127, China}

\author{Wen-Li Yang}
\affiliation{Peng Huanwu Center for Fundamental Theory, Xi'an 710127, China}
\affiliation{Insititute of Physics, Northwest University, Xi'an 710127, China}

\maketitle

In this supplementary material, we will show more details on the exact geometric mapping between Landau-Lifshitz-Gilbert and generalized nonlinear Schr\"odinger equation, and calculation details of solving the magnetic solitons.

\setcounter{equation}{0}
\renewcommand\theequation{A.\arabic{equation}}
\renewcommand\thefigure{\Roman{figure}}

\subsection{A. Exact mapping between LLG and GNLS equation: Geometric Representation}
We identify the magnetization state of ferromagnetic nanowire at any instant of time with a moving space curve in Euclidean three-dimensional space $\mathbb{E}^{3}$.
This is achieved by mapping the unit magnetization vector $\textbf{m}(x, t)$ on the unit tangent vector $\mathbf{e}_{1}$ associated with the curve. Thus the dimensionless STT-LLG equation (in the absence of damping) becomes
\begin{equation}\label{rLLG}
\begin{aligned}
\mathbf{e}_{1t}=\mathbf{e}_{1}\times\mathbf{e}_{1xx}+\mathbb{Q}\ \mathbf{e}_{1x}.
\end{aligned}
\end{equation}
In the usual way, the normal and binormal vectors of the moving space curve are constructed by taking $\mathbf{e}_{2}$ in the direction of $\mathbf{e'}_{1}$ and $\mathbf{e}_{3}=\mathbf{e}_{1}\times\mathbf{e}_{2}$.
The spatial variations of these orthogonal unit vectors is determined by the Serret-Frenet equations
\begin{equation}\label{SF}
\begin{bmatrix}
\mathbf{e}_{1} \\
\mathbf{e}_{2} \\
\mathbf{e}_{3} \\
\end{bmatrix}_{x}
=
\begin{bmatrix}
0 & \kappa & 0 \\
-\kappa & 0 & \tau\\
0 & -\tau & 0 \\
\end{bmatrix}
\begin{bmatrix}
\mathbf{e}_{1} \\
\mathbf{e}_{2} \\
\mathbf{e}_{3} \\
\end{bmatrix},
\end{equation}
where $\kappa(x,t)$ and $\tau(x,t)$ are the curvature and torsion of the space curve.
In view of (\ref{rLLG}) and (\ref{SF}) alongside the orthogonality of the three unit vectors, it is easy to obtain
\begin{equation}\label{et}
\begin{bmatrix}
\mathbf{e}_{1} \\
\mathbf{e}_{2} \\
\mathbf{e}_{3} \\
\end{bmatrix}_{t}
=
\begin{bmatrix}
0 & -\kappa\tau+\mathbb{Q}\kappa & \kappa_{x} \\
\kappa\tau-\mathbb{Q}\kappa & 0 & -\tau^{2}+\mathbb{Q}\tau+\kappa^{-1}\kappa_{xx}\\
-\kappa_{x} & \tau^{2}-\mathbb{Q}\tau-\kappa^{-1}\kappa_{xx} & 0 \\
\end{bmatrix}
\begin{bmatrix}
\mathbf{e}_{1} \\
\mathbf{e}_{2} \\
\mathbf{e}_{3} \\
\end{bmatrix}.
\end{equation}
The compatibility conditions $\frac{\partial}{\partial t}\left(\frac{\partial \mathbf{e}_{i}}{\partial x}\right)=\frac{\partial}{\partial x}\left(\frac{\partial \mathbf{e}_{i}}{\partial t}\right),\ i=1,2,3 $, between Eqs. (\ref{SF}) and (\ref{et}) lead to the following evolution equations for $\kappa$ and $\tau$
\begin{subequations}\label{kt}
\begin{align}
&\kappa_{t}=-\left(\kappa\tau_{x}+2\kappa_{x}\tau\right)-\mathbb{Q}\kappa_{x}, \label{kt1} \\
&\tau_{t}=\left(\kappa^{-1}\kappa_{xx}-\tau^{2}\right)_{x}+\kappa\kappa_{x}-\mathbb{Q}\tau_{x}. \label{kt2}
\end{align}
\end{subequations}
On making the complex transformation
$\Psi=\frac{1}{2}\kappa\exp\left(i\int\tau{\rm d}x\right)$,
we finally arrive at a generalized nonlinear Schr\"odinger (GNLS) equation (it is easy to verify that the real and imaginary parts of (\ref{GNLS}) is equivalent to (\ref{kt1}) and (\ref{kt2}), respectively)
\begin{equation}\label{GNLS}
i\Psi_{\tau}+\Psi_{\zeta\zeta}+2|\Psi|^{2}\Psi-i \mathbb{Q} \Psi_{\zeta}=0.
\end{equation}
Thus we have proved that the STT-LLG equation can be exactly mapped into the integrable GNLS equation.

\setcounter{equation}{0}
\renewcommand\theequation{B.\arabic{equation}}
\renewcommand\thefigure{\Roman{figure}}

\subsection{B. Lax Representation and Darboux Transformation}
We now turn to establish the connection between the solutions of the LLG equation and the GNLS equation.
Using the Pauli matrices ($\sigma_{1}, \sigma_{2}, \sigma_{3}$), the LLG equation can be rewriten into the matrix form
\begin{equation}\label{VLLG}
\begin{aligned}
\mathbf{\widehat{m}}_{t}=\frac{1}{2i}\left[\mathbf{\widehat{m}}, \mathbf{\widehat{m}}_{xx}\right]+\mathbb{Q}\ \mathbf{\widehat{m}}_{x},
\end{aligned}
\end{equation}
where
$\mathbf{\widehat{m}}=m^{x}\sigma_{1}+m^{y}\sigma_{2}+m^{z}\sigma_{3}$ and $\left[\cdot , \cdot\right]$ denotes the Lie bracket of the matrices. For this equation, the boundary condition is given by $\lim\limits_{x\to\pm\infty}\mathbf{\widehat{m}}=\sigma_{3}$, i.e., $\lim\limits_{x\to\pm\infty}\mathbf{m}=(0,0,1)$.
Considering the Lax representation of the GNLS (\ref{GNLS})
\begin{equation}\label{Lax1}
\frac{\partial \Phi}{\partial x}={\mathbf U} \Phi,\quad \frac{\partial \Phi}{\partial t}={\mathbf V} \Phi,
\end{equation}
where ${\mathbf U}={\mathbf U}_{0}+\lambda{\mathbf U}_{1}, {\mathbf V}={\mathbf V}_{0}+\lambda{\mathbf V}_{1}+\lambda^{2}{\mathbf V}_{2}$ and $\lambda$ is the spectral parameter,
\begin{equation}\label{UV1}
{\mathbf U}_{0}=
\begin{pmatrix}
0 &  \Psi \\
-\Psi^{*} & 0 \\
\end{pmatrix},\quad
{\mathbf U}_{1}=-\sigma_{3},\quad
{\mathbf V}_{0}=
\begin{pmatrix}
i\lvert \Psi \rvert^{2} &  i\Psi_{x}+\mathbb{Q} \Psi \\
i\Psi^{*}_{x}-\mathbb{Q} \Psi^{*} & -i\lvert \Psi \rvert^{2} \\
\end{pmatrix},\quad
{\mathbf V}_{1}=-2i{\mathbf U}_{0}-\mathbb{Q}\sigma_{3},\quad
{\mathbf V}_{2}=2i\sigma_{3}.
\end{equation}
Suppose $\Phi_{1}(x,t,\lambda)$ and $\Phi_{2}(x,t,\lambda)$ are two linear independence eigenvectors of Lax pair (\ref{Lax1}), then $\Omega=\left(\Phi_{1},\Phi_{2}\right)$ also satisfies Eq. (\ref{Lax1}).
Let $g(x,t)=\Omega(x,t,\lambda) \big|_{\lambda=0}$, we have $g_{x}={\mathbf U}_{0}g,\ g_{t}={\mathbf V}_{0}g$. From transformation $\widehat{\Phi}=g^{-1}\Phi$, we obtain
\begin{equation}\label{Lax2}
\frac{\partial \widehat{\Phi}}{\partial x}=\widehat{\mathbf U} \widehat{\Phi},\quad \frac{\partial \widehat{\Phi}}{\partial t}=\widehat{\mathbf V} \widehat{\Phi},
\end{equation}
where
\begin{equation}\label{UV2}
\begin{aligned}
&\widehat{\mathbf U}=g^{-1}{\mathbf U}g-g^{-1}g_{x}=g^{-1}({\mathbf U}-{\mathbf U}_{0})g=\lambda g^{-1}{\mathbf U}_{1}g,\\
&\widehat{\mathbf V}=g^{-1}{\mathbf V}g-g^{-1}g_{t}=g^{-1}({\mathbf V}-{\mathbf V}_{0})g=\lambda g^{-1}{\mathbf V}_{1}g+\lambda^{2} g^{-1}{\mathbf V}_{2}g,
\end{aligned}
\end{equation}
Let $\mathbf{\widehat{m}}=-g^{-1}\sigma_{3}g$ be a solution of Eq. (\ref{VLLG}), then
\begin{equation}\label{mmx}
\mathbf{\widehat{m}}\mathbf{\widehat{m}}_{x}=-g^{-1}\sigma_{3}{\mathbf U}_{0}\sigma_{3}g+g^{-1}\sigma_{3}^{2}{\mathbf U}_{0}g=2g^{-1}{\mathbf U}_{0}g,
\end{equation}
Substitute (\ref{UV1}) into (\ref{UV2}), together with the definition of $\mathbf{\widehat{m}}$ and (\ref{mmx}), matrices (\ref{UV2}) of new Lax pair (\ref{Lax1}) can be rewritten as
\begin{equation}\label{UV3}
\begin{aligned}
&\widehat{\mathbf U}=\lambda\mathbf{\widehat{m}},
&\widehat{\mathbf V}=\lambda\mathbb{Q}\mathbf{\widehat{m}}-i\lambda\mathbf{\widehat{m}}\mathbf{\widehat{m}}_{x}-2i\lambda^{2}\mathbf{\widehat{m}},
\end{aligned}
\end{equation}
Using the factor $\mathbf{\widehat{m}}^{2}=\mathbf{I}$, the compatibility condition $\widehat{\mathbf U}_{t}-\widehat{\mathbf V}_{x}+[\widehat{\mathbf U},\widehat{\mathbf V}]=0$ exactly yields the matrix form LLG equation (\ref{VLLG}).
Thus we proved the Lax gauge equivalence of the GNLS equation (\ref{GNLS}) and the dimensionless STT-LLG equation.
Through the established gauge equivalence detailed above, it becomes evident that given a non-zero solution $\Phi$ of the GNLS equation (\ref{GNLS}), the corresponding eigenfunctions can be derived via Lax pair (\ref{Lax1}). This process thereby elucidates the determination of the invertible matrices $\Omega(x,t,\lambda)$ and $g(x,t)$. Further through the transformation $\widehat{\Phi}=g^{-1}\Phi$ and $\mathbf{\widehat{m}}=-g^{-1}\sigma_{3}g$, we are enabled to acquire the solution $\mathbf{\widehat{m}}$ for (\ref{VLLG}). Finally, the three components of magnetization, namely $m^{x}$, $m^{y}$, and $m^{z}$, can be obtained from the definition of $\mathbf{\widehat{m}}$, constituting the non-trivial solution to the original STT-LLG equation.

To obtain the dynamical magnetic soliton in the ferromagnetic nanowire, we are going to construct the Darboux transformation of (\ref{GNLS}).
Let $\Phi_{1}^{[0]}(x,t,\lambda)$ and $\Phi_{2}^{[0]}(x,t,\lambda)$ be the eigenfunction of the Lax pair (\ref{Lax1}) corresponding to the zero solution of the GNLS equation (\ref{GNLS}).
Demonstrating the reciprocity of the Lax pair solution with respect to spectral parameters, it is straightforward to establish that if $\big(\Phi_{1}^{[0]}(x,t,\lambda_{1}), \Phi_{2}^{[0]}(x,t,\lambda_{1})\big)^{T}$ represents the solution for Lax pair (\ref{Lax1}) corresponding to the spectral parameter $\lambda_1$, then $\big(\Phi_{2}^{[0]*}(x,t,\lambda_{1}^{*}), -\Phi_{1}^{[0]*}(x,t,\lambda_{1}^{*})\big)^{T}$ constitutes the solution for the corresponding spectral parameter $\lambda_1^{*}$. Denote
\begin{equation}\label{H}
\mathbf{H}_{1}=
\begin{pmatrix}
\Phi_{1}^{[0]} & \Phi_{2}^{[0]*} \\
\Phi_{2}^{[0]} & -\Phi_{1}^{[0]*} \\
\end{pmatrix},\quad
\Lambda_{1}=
\begin{pmatrix}
\lambda_{1} & 0 \\
0 & \lambda_{1}^{*} \\
\end{pmatrix},
\end{equation}
where $\big(\Phi_{1}^{[0]},\Phi_{2}^{[0]}\big)^{T}=\big(\exp[-\lambda_{1}x+2i\lambda_{1}^{2}t-\lambda_{1}\int\mathbb{Q}\ {\rm d}t],\exp[\lambda_{1}x-2i\lambda_{1}^{2}t+\lambda_{1}\int\mathbb{Q}\ {\rm d}t]\big)^{T}$.
The Darboux matrix is acquired through the standard procedure
\begin{equation}\label{T}
\mathbf{T}^{[1]}=\lambda\mathbf{I}-\mathbf{H}_{1}\Lambda_{1}\mathbf{H}_{1}^{-1},
\end{equation}
leading to the solution $\Phi^{[1]}=\mathbf{T}^{[1]}\Phi$ of new spectral problem.
Therefore the Darboux transformation is written as
\begin{equation}\label{DT}
\Psi^{[1]}(x,t)=\Psi^{[0]}(x,t)-\frac{2(\lambda_{1}+\lambda_{1}^{*})\Phi_{1}^{[0]}\Phi_{2}^{[0]*}}{\lvert \Phi_{1}^{[0]} \rvert^{2}+\lvert \Phi_{2}^{[0]} \rvert^{2}}.
\end{equation}
Taking $\lambda_{1}=a+ib$ we get the soliton solution of GNLS equation
\begin{equation}\label{Sol}
\Psi^{[1]}(x,t)=-2a\sech\left[2a(x+4bt+\mathbb{Q}t)\right]\exp\left[2i\left[(2a^{2}-2b^{2}-\mathbb{Q})t-bx\right]\right],
\end{equation}
and the corresponding eigenfunction $\Phi^{[1]}(x,t,\lambda)=(\Phi_{1}^{[1]},\Phi_{2}^{[1]})^{T}$.
Substitute the above results into $\Omega(x,t,\lambda)=\begin{pmatrix}\Phi_{1}^{[1]} & \Phi_{2}^{[1]*} \\ \Phi_{2}^{[1]} & -\Phi_{1}^{[1]*} \end{pmatrix}$, $g(x,t)=\Omega(x,t,\lambda)\big|_{\lambda=0}$, $\mathbf{\widehat{m}}=-g^{-1}\sigma_{3}g$ in sequence, we finally obtain exact cycloidal chiral
magnetic soliton solution in ferromagnetic nanowires under the influence of spin current injection

\begin{equation}\label{One}
\begin{split}
m^{x}&=\frac{\mathbf{\widehat{m}}_{12}+\mathbf{\widehat{m}}_{21}}{2}=\frac{2a}{a^{2}+b^{2}}\left[a\sinh(\Xi)\sin(\Gamma)-b\cosh(\Xi)\cos(\Gamma)\right]\cdot\sech^{2}(\Xi),\\
m^{y}&=\frac{\mathbf{\widehat{m}}_{21}-\mathbf{\widehat{m}}_{12}}{2i}=\frac{2a}{a^{2}+b^{2}}\left[a\sinh(\Xi)\cos(\Gamma)+b\cosh(\Xi)\sin(\Gamma)\right]\cdot\sech^{2}(\Xi),\\
m^{z}&=\mathbf{\widehat{m}}_{11}=1-\frac{2a^{2}}{a^{2}+b^{2}}\sech^{2}(\Xi),\\
\end{split}
\end{equation}
with
\begin{equation*}
\Xi=2a\left[x+\int(\mathbb{Q}+4 b){\rm d}t\right],\quad
\Gamma=2b\left[x+\int(\mathbb{Q}-2(a^{2}-b^{2})/b){\rm d}t\right],
\end{equation*}

\noindent where $a$ and $b$ describe the wave number and the velocity of the chiral soliton.

\begin{figure}[htb]
\subfigure[]{
\begin{minipage}[t]{0.45\linewidth}
\centering
\includegraphics[width=0.9\textwidth]{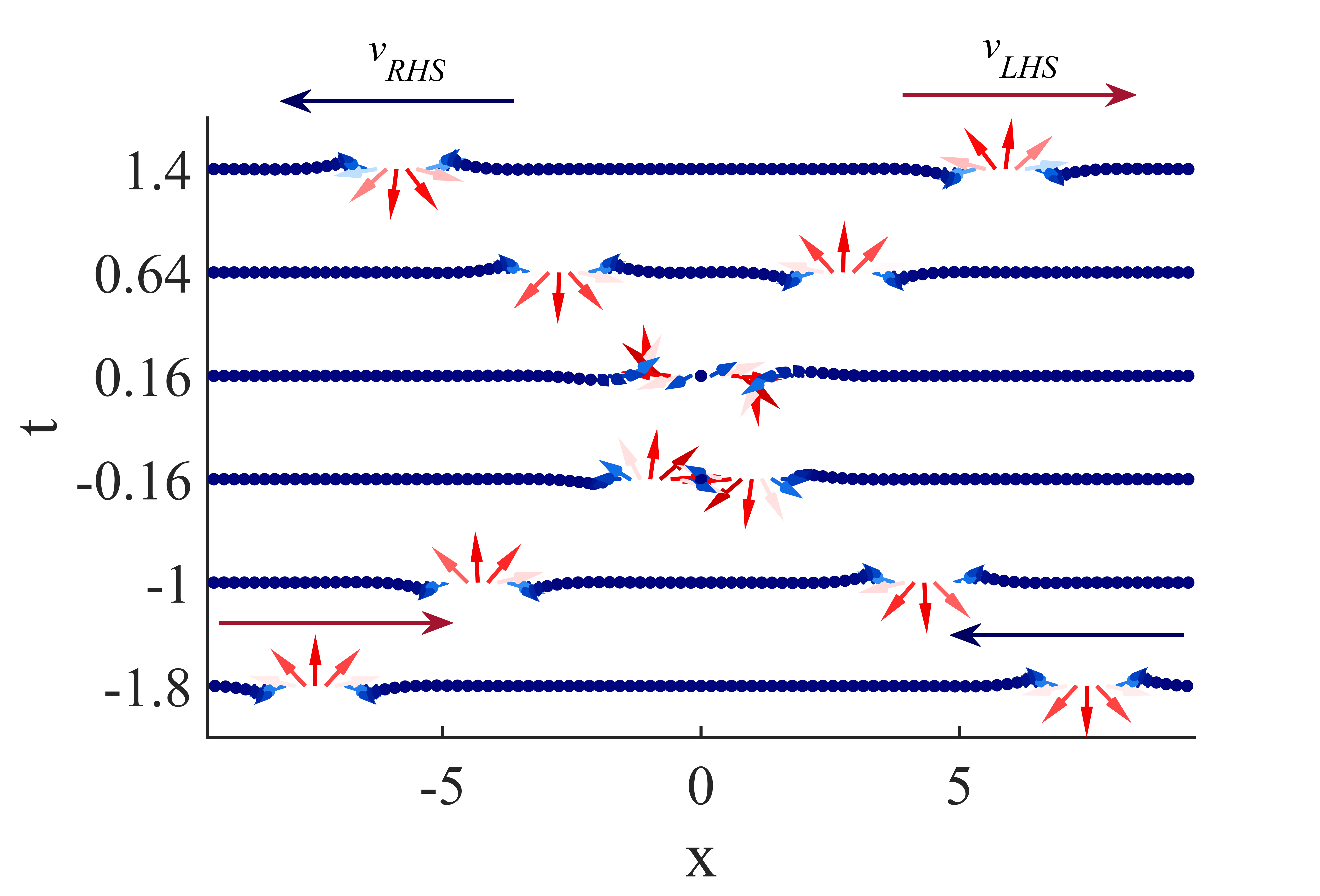}
\end{minipage}
}
\subfigure[]{
\begin{minipage}[t]{0.45\linewidth}
\centering
\includegraphics[width=0.9\textwidth]{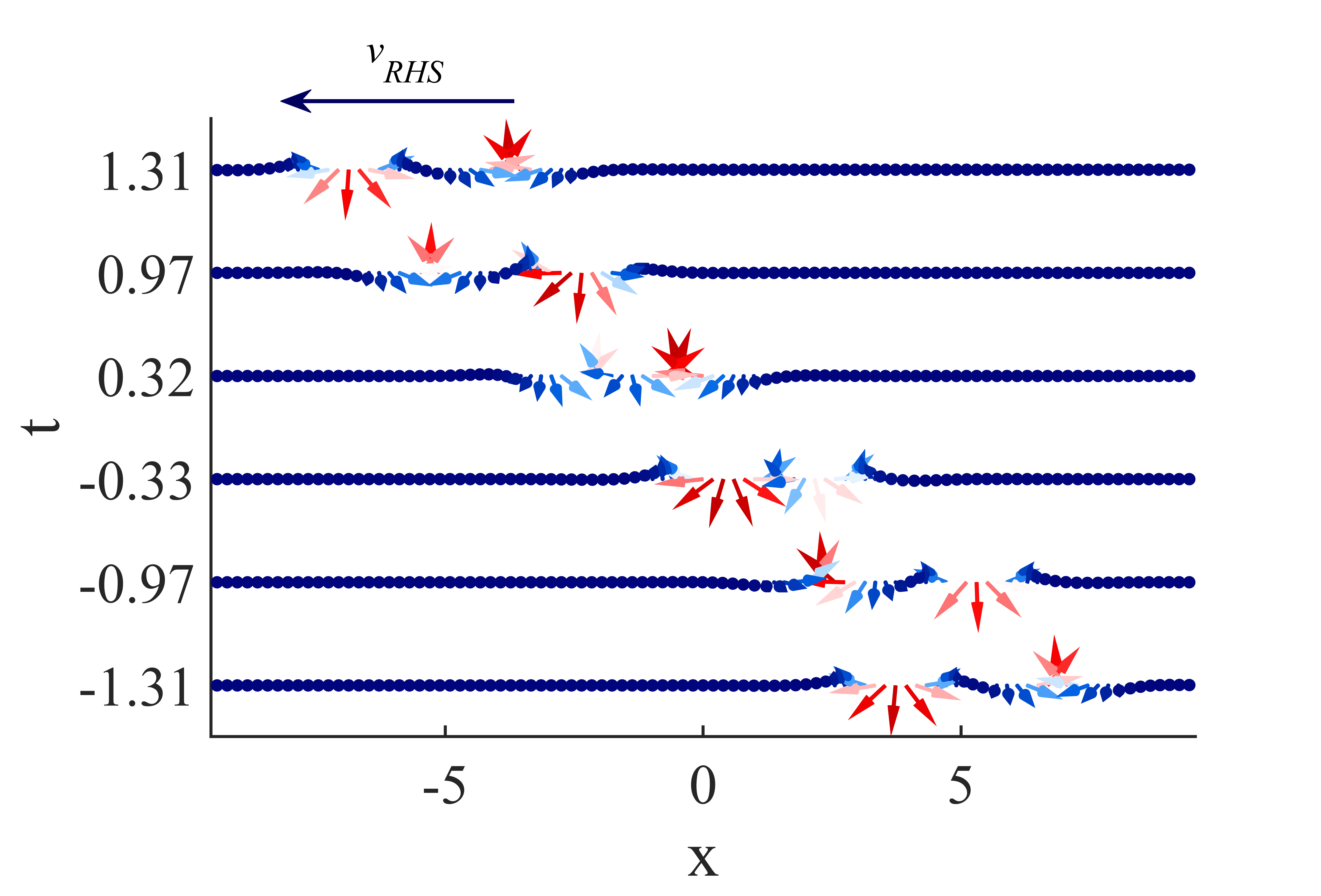}
\end{minipage}
}
\caption{The interaction between two chiral magnetic solitons. (a) Interaction between left-handed and right-handed magnetic soliton. (b) Bound states formed by two right-handed magnetic Solitons.}\label{ex}
\end{figure}

To gain deeper insight into the interaction dynamics between two chiral
magnetic solitons, we continue to utilize gauge transformation (2.2.67) to construct two-soliton solutions based on the above single soliton solutions.
The second-order Darboux matrix is expressed as follows
\begin{equation}\label{HLT2}
\mathbf{H}_{2}=
\begin{pmatrix}
\Phi_{1}^{[1]} & \Phi_{2}^{[1]*} \\
\Phi_{2}^{[1]} & -\Phi_{1}^{[1]*} \\
\end{pmatrix},\quad
\Lambda_{2}=
\begin{pmatrix}
\lambda_{2} & 0 \\
0 & \lambda_{2}^{*} \\
\end{pmatrix},\quad
\mathbf{T}^{[2]}=\lambda\mathbf{I}-\mathbf{H}_{2}\Lambda_{2}\mathbf{H}_{2}^{-1},
\end{equation}
and the specific Darboux transformation form of the two-soliton solution is subsequently obtained
\begin{equation}\label{DT2}
\Psi^{[2]}(x,t)=\Psi^{[1]}(x,t)-\frac{2(\lambda_{2}+\lambda_{2}^{*})\Phi_{1}^{[1]}\Phi_{2}^{[1]*}}{\lvert \Phi_{1}^{[1]} \rvert^{2}+\lvert \Phi_{2}^{[1]} \rvert^{2}}.
\end{equation}
Taking $\lambda_{1}=a_{1}+ib_{1}, \lambda_{2}=a_{2}+ib_{2}$, after tedious simplification, we get two-soliton solution of GNLS equation

\begin{equation}\label{Sol2}
\Psi^{[2]}(x,t)=4\frac{\eta_{1}e^{i\beta_{2}}\cosh(\alpha_{2})+\eta_{2}e^{i\beta_{1}}\cosh(\alpha_{1})+i\eta_{3}\left(e^{i\beta_{1}}\sinh(\alpha_{1})-e^{i\beta_{2}}\sinh(\alpha_{2})\right)}
{\eta_{4}\cosh(\alpha_{1}+\alpha_{2})+\eta_{5}\cosh(\alpha_{1}-\alpha_{2})+\eta_{6}\cos(\beta_{1}-\beta_{2})},
\end{equation}
where
\begin{equation*}
\begin{split}
&\alpha_{1}=2a_{1}(x+4b_{1}t+\int\mathbb{Q}\ {\rm d}t),\quad
\beta_{1}=4(a^{2}-b^{2})t-2b_{2}(x+\int\mathbb{Q}\ {\rm d}t),\quad \\
&\alpha_{2}=2a_{2}(x+4b_{2}t+\int\mathbb{Q}\ {\rm d}t),\quad \beta_{2}=4(a^{2}-b^{2})t-2b_{2}(x+\int\mathbb{Q}\ {\rm d}t),\quad \\
&\eta_{1}=[(a_{2}^{2}-a_{1}^{2})-(b_{2}-b_{1})^{2}]a_{1},\quad \eta_{4}=-(a_{2}-a_{1})^{2}-(b_{2}-b_{1})^{2},\quad
\eta_{3}=2a_{1}a_{2}(b_{2}-b_{1}),\quad \\
&\eta_{2}=[(a_{1}^{2}-a_{2}^{2})-(b_{1}-b_{2})^{2}]a_{2},\quad \eta_{5}=-(a_{2}+a_{1})^{2}-(b_{2}-b_{1})^{2},\quad
\eta_{6}=4a_{1}a_{2}. \quad \\
\end{split}
\end{equation*}
Continuing with the same approach in the previous text, we are able to provide a precise expression for the three-component of the magnetization $m$ for the dynamic chiral magnetic two-solitons. Owing to the complexity of its explicit expression, we opt to omit it and solely showcase the corresponding figure.
Two typical solutions for the interaction between two chiral magnetic solitons are shown in Fig.(\ref{ex}).

\setcounter{equation}{0}
\renewcommand\theequation{C.\arabic{equation}}
\renewcommand\thefigure{\Roman{figure}}

\subsection{C. Chirality of cycloidal chiral magnetic soliton in Bloch sphere}

As shown in the main text, we denote the polar and azimuthal angles of the vector $\textbf{m}$ as $\theta$ and $\varphi$, respectively. This notation allows us to express $M_{+}=m^{x}+im^{y}=M_{0}\sin(\theta)\exp(i\varphi)$ and $m^{z}=M_{0}\cos(\theta)$. By employing the three-component analytical formulation, we can infer the inverse solution for $\theta$ and $\varphi$, which in turn can be mapped onto the Bloch unit sphere. This approach yields a trajectory map delineating the movement of chiral magnetic solitons across the unit sphere.
Consequently, within the magnetization unit sphere, a chiral magnetic soliton traces a closed curve encompassing a single pole. The trajectories of motion for the two distinct types of chiral magnetic soliton solutions on the Bloch spheres can indirectly manifest their chirality. Commencing from negative infinity, which corresponds to the pole of ground state, the left and right-handed chiral magnetic solitons will give rise to enclosed paths, one proceeding in a clockwise direction and the other counterclockwise. This motion pattern eventually in mirror-symmetrical trajectories.

\begin{figure}[h]
\centering
\includegraphics[width=0.65\textwidth]{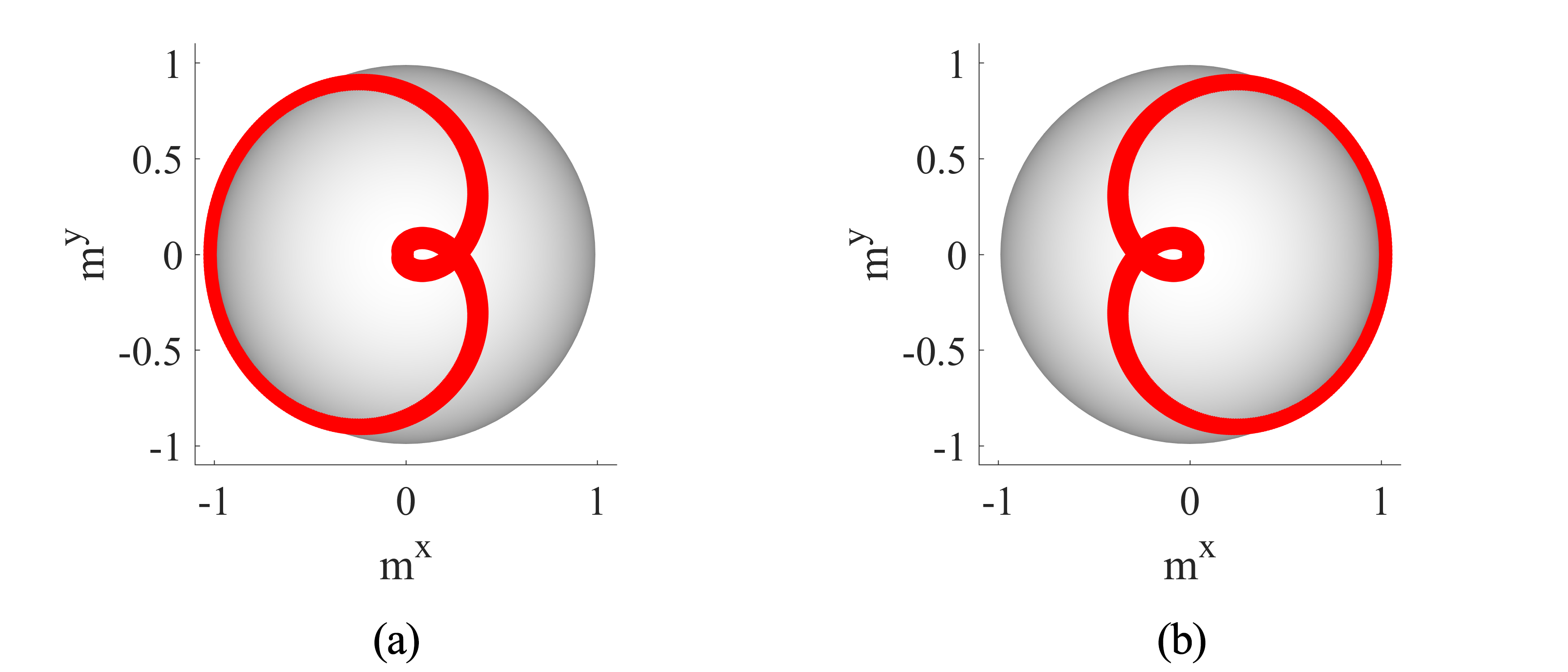}
\caption{Trajectories of chiral magnetic solitons on the Bloch sphere at time t=0. (a) Left-handed chiral magnetic soliton $a=1$, $b=1$, (b) Right-handed chiral magnetic soliton $a=1$, $b=-1$.}\label{Sketch}
\end{figure}

\bibliographystyle{unsrt}
\bibliography{references}